\documentclass[a4paper,10pt,aps,amsmath,floatfix,longbibliography,prb,notitlepage,twocolumn
,superscriptaddress,showkeys]{revtex4-1}
\usepackage{graphicx} 
\usepackage{amsmath}
\usepackage{enumerate} 
\usepackage[caption=false]{subfig}
\usepackage[colorlinks=true,linkcolor=red]{hyperref}
\usepackage{xcolor}
\usepackage{hyperref}
\newcommand{\be}{\begin{equation}} 
\newcommand{\ee}{\end{equation}} 

\begin{document}
\title{Equilibrium via multi-spin-flip Glauber dynamics in Ising Model}
\author{Diana Thongjaomayum} \affiliation{Department of Physics, Tezpur 
University, Assam} \author{Prabodh 
Shukla} \affiliation{Retired, North-Eastern Hill University, Shillong}
\date{\today}

\begin{abstract} 
Notwithstanding great strides that statistical mechanics 
has made in recent decades, an analytic solution of arguably the 
simplest model of relaxation dynamics, the Ising model in an applied 
external field remains elusive even in $1d$. Extant studies are based on 
numerics using single-spin-flip Glauber dynamics. There is no reason why 
this algorithm should lead to the global minimum energy state of the 
system. With this in mind, we explore multi-spin-flip parallel and 
sequential Glauber dynamics of Ising spins in $1d$ and also on a regular 
random graph of coordination number $z=3$. We view our study as a small 
initial step to test the generally implied hypothesis that the 
equilibrium is independent of the relaxational dynamics or if it carries 
some signature of it.
\end{abstract} 


\maketitle

\section{Introduction}

The grand task of statistical mechanics is to explain thermodynamics of 
systems with huge degrees of freedom. It is evidently a challenging task 
but made relatively easy in the domain of equilibrium statistical 
mechanics. A large class of isolated systems when left to themselves 
settle down in an equilibrium state which is characterized by small 
fluctuations with average value equal to zero on practical time scales. 
We can ignore such fluctuations and bypass their complicated dynamics. A 
simple overriding principle that a system in equilibrium is in a state 
of minimum free energy suffices to explain the observed behavior. Phase 
transition points and critical phenomena remained a challenge in 
equilibrium statistical mechanics for a long time because fluctuations 
at the critical point occur at all length scales. However, these too 
were tamed in recent decades by new techniques of the renormalization 
group theory ~\cite{wilson}.

In contrast, in the domain of nonequilibrium statistical mechanics, the 
fluctuations in the system do not average out to zero as the system 
approaches equilibrium very slowly or evolves in an inconveniently 
sporadic manner over observed time scales. In this difficult situation 
there is evidently no recourse available to us except to solve the 
equations of motion for impossibly large number of coupled degrees of 
freedom. Even to get a grip on the general principles of nonequilibrium 
statistical mechanics we have to be content only with a caricature of the 
system and its dynamics by some toy model. The Ising model and Glauber 
dynamics fit this bill ~\cite{ising,onsager,glauber,bertotti}. In the 
following we describe our model and dynamics in detail. However it is 
not out of place here to mention the importance of hysteresis in our 
approach.

How do we know when a system has reached equilibrium or how far is it 
from it? This is not an easy question. In our approach, we rely on 
hysteresis for an answer. Hysteresis stands for history dependent 
dynamics of a system. The magnetization trajectory in increasing applied 
field is distinct from the one in decreasing field. The difference 
between the two is a measure of hysteresis in the system. The 
equilibrium state is by definition supposed to be independent of the 
initial state of the system. Thus hysteresis in the system serves as a 
signature of the distance from equilibrium in an evolving system. We may 
conclude that a system has reached equilibrium when it shows zero 
hysteresis. In our approach we consider two copies of a system 
magnetized to a saturation state in opposite directions and let the two 
copies evolve at the same value of the applied field for the same length 
of time. The difference between the two final states is taken as the 
distance of the system from equilibrium at that applied field.

\section{The Model}

The Hamiltonian of the $1d$ Ising model is, \be H = -J \sum_i s_i 
s_{i+1}-h \sum_i s_i; s_i = \pm 1 \ee \noindent Here $J$ is 
ferromagnetic ($J>0$) nearest neighbor interaction, $h$ is an applied 
field that is ramped up or down in suitable steps, $s_i$ is an Ising 
spin at site-$i$ on a chain of $N$ sites, $i=1, 2, \ldots, N$. Periodic 
boundary conditions are assumed, i.e. $s_i \equiv 
s_{i+N}$. The magnetization per site 
is defined as $m=\sum_i s_i/N$.

The stochastic single-spin-flip Glauber dynamics associated with the 
Hamiltonian is prescribed as follows. The effective local field acting 
on spin $s_i(t)$ at site-$i$ at time $t$ is equal to $f_i(t)=J 
[s_{i-1}(t) + s_{i+1}(t) ] + h $. The energy of the spin at site-$i$ is 
equal to $e_i(t)=-f_i(t) s_i(t)$ i.e. the energy is negative if $s_i(t)$ 
is aligned along the effective field $f_i(t)$, and positive otherwise. 
At each update moment the spin can flip with a probability $P_i(t)$ or 
stay unchanged with probability $1-P_i(t)$. The probability of flipping 
is larger if it lowers the energy. At temperature $T$, the probability 
of flipping is given by the following expression in terms of 
$K=J/(k_BT)$, $k_B$ being the Boltzmann constant,

\begin{align}
P_i(s_i \rightarrow -s_i) &= \frac{\exp(-K f_i s_i)}{\exp(-K f_is_i) 
+ \exp(K f_i s_i)} \nonumber \\ &=\frac{1}{1+\exp(2K e_i)}
\end{align}

\noindent Under parallel dynamics the entire set of spins $\{s_i(t)\}$ 
at time-$t$ is updated simultaneously producing $m(t+1)$ from $m(t)$ in 
one Monte Carlo time step or one MC cycle. Under sequential dynamics the 
spins are updated one at a time sequentially starting at site-$1$ and 
ending at site-$N$. For meaningful comparison with parallel dynamics we 
consider the update of the entire chain $i=1,\ldots,N$ in sequential 
dynamics as one MC cycle. We now present numerical results setting $J=1$ 
and $K_B=1$ for convenience, so that $K$ may be thought of as inverse 
temperature.

\section{Simulations}

\subsection{single-spin-flip dynamics}

\begin{figure}[hbt]
 \centering

\includegraphics[width=0.45\textwidth,angle=0]{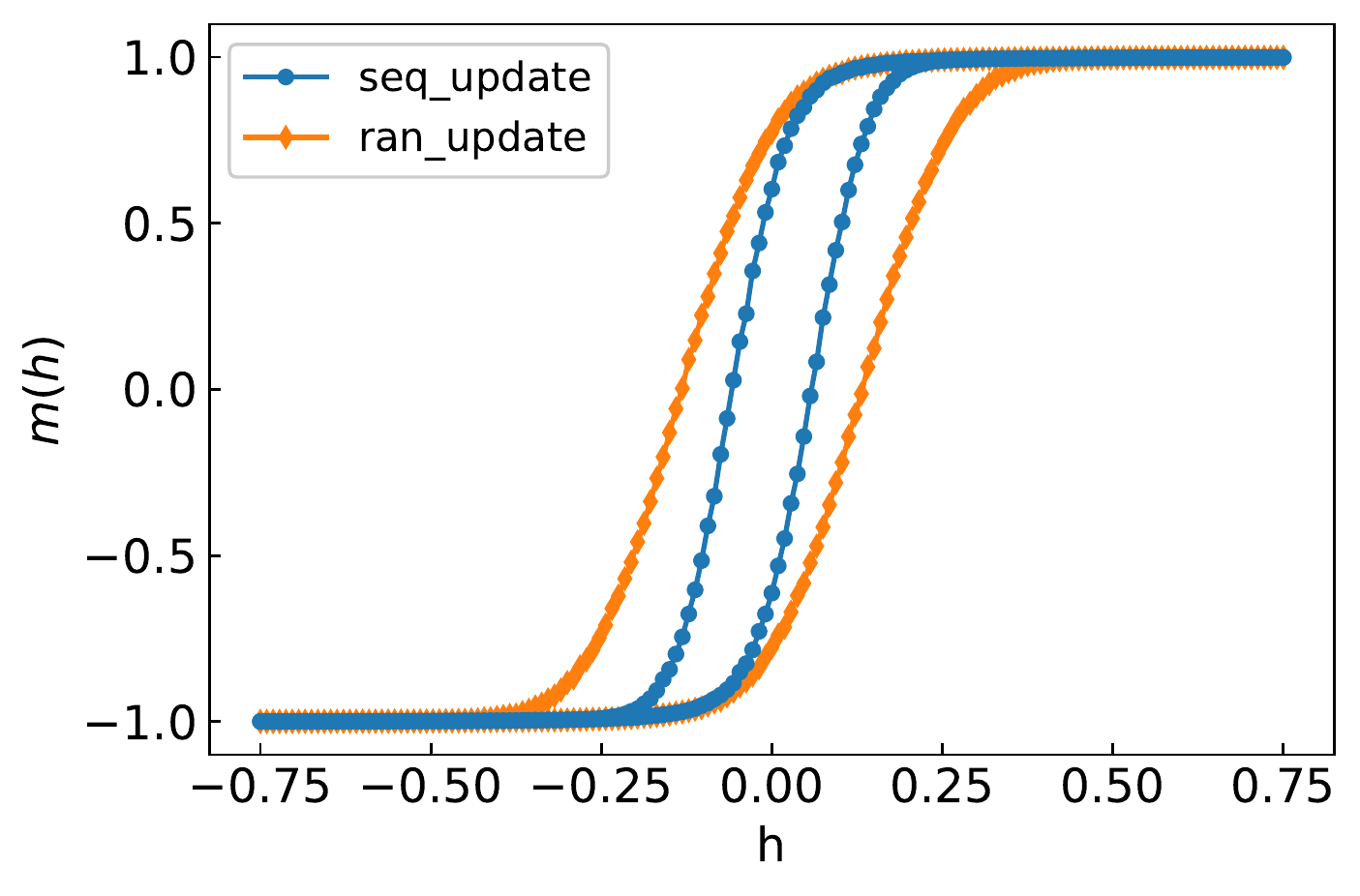}
   \caption{(Color online) Comparison for 1 spin-flip random vs 
   sequential updating of spins: $L=2000$, $K=1.5$ 
   and $t=50$ averaged over 200 configurations. }
  \label{fig:1}
\end{figure}  

Fig.1 shows two hysteresis loops using single-spin-flip Glauber 
dynamics. A narrow loop obtained by sequential dynamics is symmetrically 
nested inside a larger loop for parallel dynamics. Both loops are for 
$K=1.5$ on a chain of length $N=2000$. At each applied field $h$, the 
system is allowed to evolve for $t=50$ MC cycles from an initial state 
with all spins parallel to each other. For the lower half of the loop, 
the initial state has all spins down and the field is ramped up in 
suitably small steps. On the upper half of the loop we start with all 
spins up and decrease the field till they are all turned down.  The 
incremental steps for $h$ are chosen sufficiently small so that the data 
points for corresponding magnetization $m(h)$ appear as a smooth curve 
to the eye. Also contributing to the apparent smoothness of the curves 
is that each data point is an average over $200$ runs of the stochastic 
dynamics. We note that the hysteresis curves look remarkable similar to 
the observed hysteresis curves in the laboratory despite the model being 
just a toy model. Presumably this is the effect of the Boltzmann factor 
used in the Glauber dynamics. The Boltzmann factor determines thermal 
relaxation in any natural system. Why does the sequential dynamics yield 
a narrower loop? This is not too difficult to understand on reflection. 
In sequential dynamics each spin is relaxed in the shadow of its 
neighbor that has been relaxed earlier. So effectively the system has 
more opportunity to lower its energy in one MC cycle under sequential 
dynamics than under parallel dynamics. We are primarily interested in 
exploring algorithms that are more efficient for thermal relaxation. 
Therefore we confine ourselves to sequential dynamics in the 
following presentation. Of course the order in which $N$ spins are chosen 
for updating can be random rather than prefixed. We may choose the first 
spin randomly and then update all other spins sequentially along the 
chain. Or we may choose each of $N$ spins in one MC cycle randomly. We 
have explored this freedom to some extent but it does not seem to make a 
great deal of difference.

As may be expected, the hysteresis loops shown in Fig.1 become narrower 
when the system is relaxed for longer time. At $t=1000$ MC cycles the 
lower and upper halves of the loop appear to merge with each other on 
the scale of the figure. We interpret it to mean that the system has 
reached equilibrium. The equilibrium magnetization in an Ising chain in 
an external field $h$ can be determined analytically. Fig.2 shows that 
the equilibrium magnetization curve is indistinguishable from the 
collapsed hysteresis loops. This figure also shows the results of a 
simulation based on a two-spin-flip process which we discuss below. 
Suffice it to note here that the curve for the two-spin process at a 
lower value of $t=500$ is already indistinguishable from the 
single-spin-flip case at $t=1000$. Although the magnetization curves for 
one and two-spin flips in the long time limit merge with the equilibrium 
curve, it would be wrong to conclude that the corresponding 
thermodynamic states are identical. The equilibrium magnetization is 
based on a state which is obtained from the partition function. It takes 
into account all fluctuations in the system with corresponding Boltzmann 
weights. The long time limits obtained from dynamics necessarily 
comprise a restricted class of fluctuations. The correct conclusion to 
draw from our numerical results so far is that the average magnetization 
is not able to resolve the fine differences that must exist. We must 
search for other more revealing signatures of the differences.

\subsection{two-spin-flip dynamics}

There is no unique or standard algorithm to extend the single-spin-flip 
Glauber dynamics to a more general relaxation process. A rather obvious 
path is to incorporate simultaneous flips of a pair of neighboring spins 
or larger cluster of spins. Even here many variants are possible. We 
choose arguably the simplest of possible procedures which is as follows. 
Each MC cycle consists of two sub-cycles. The first sub-cycle implements 
single-spin-flips, and the next pairs of neighboring spins. The results 
are qualitatively similar to what might be expected from two successive 
MC cycles of single-spin-flips but are not identical to these. Fig.3 
illustrates the result of a simulation for the same choice of parameters 
as in the earlier figures. As may be expected, two-flips move the system 
closer to equilibrium and consequently the loop narrows significantly.

Fig.1, Fig.2, and Fig.3 show hysteresis loops at a single value of 
inverse temperature $K=1.5$ but obtained by different procedures as 
described above. A remark about qualitative change in the shape of loop 
with varying $K$ is in order. At lower values of $K$ and moderate time 
periods $t$, the loop tends to elongate along the $x-$axis i.e. there is 
significant separation between lower and the upper halves over a larger 
range of applied field and remnant magnetization is relatively small\cite{shukla1}. 
This is understandable because higher thermal energy weakens the 
ordering tendency of the applied field at moderate values of the applied 
field. At lower temperature and correspondingly higher $K$, the loop is 
elongated along the $y-$axis. In principle, hysteresis should vanish in 
the limit $t \rightarrow \infty$ irrespective of the value of $K$. 
However the relaxation becomes extremely slow at very large $K$ and it 
is difficult to observe the equilibrium limit of magnetization 
approaching a step function jump from $m=-1$ to $m=+1$ at $h=0$.

\begin{figure}[htb]
 \centering
\includegraphics[width=0.45\textwidth,angle=0]{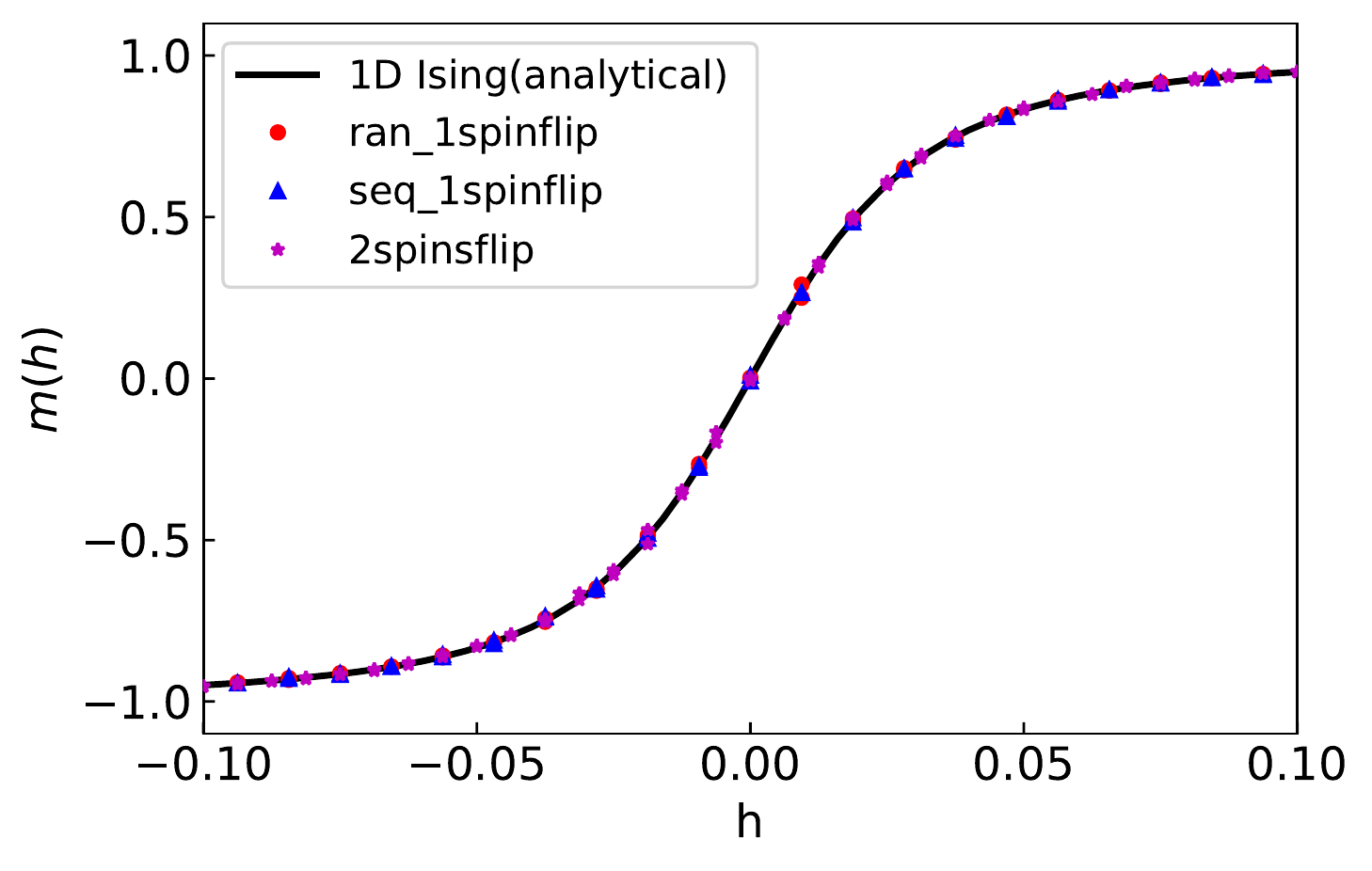}

   \caption{(Color online) Figure shows the convergence of the 
   hysteresis loops of Fig1 to the equilibrium $1d$ Ising analytical 
   values on increasing the relaxation time, $t=1000$. All data are 
   averaged over 100 configurations. The curve corresponding to 
   2-spin-flip for $L=2000$, $K=1.5$ but $t=500$ is also shown for 
   comparison. }
  \label{fig:2}
\end{figure}  

\begin{figure}[htb]
 \centering
\includegraphics[width=0.45\textwidth,angle=0]{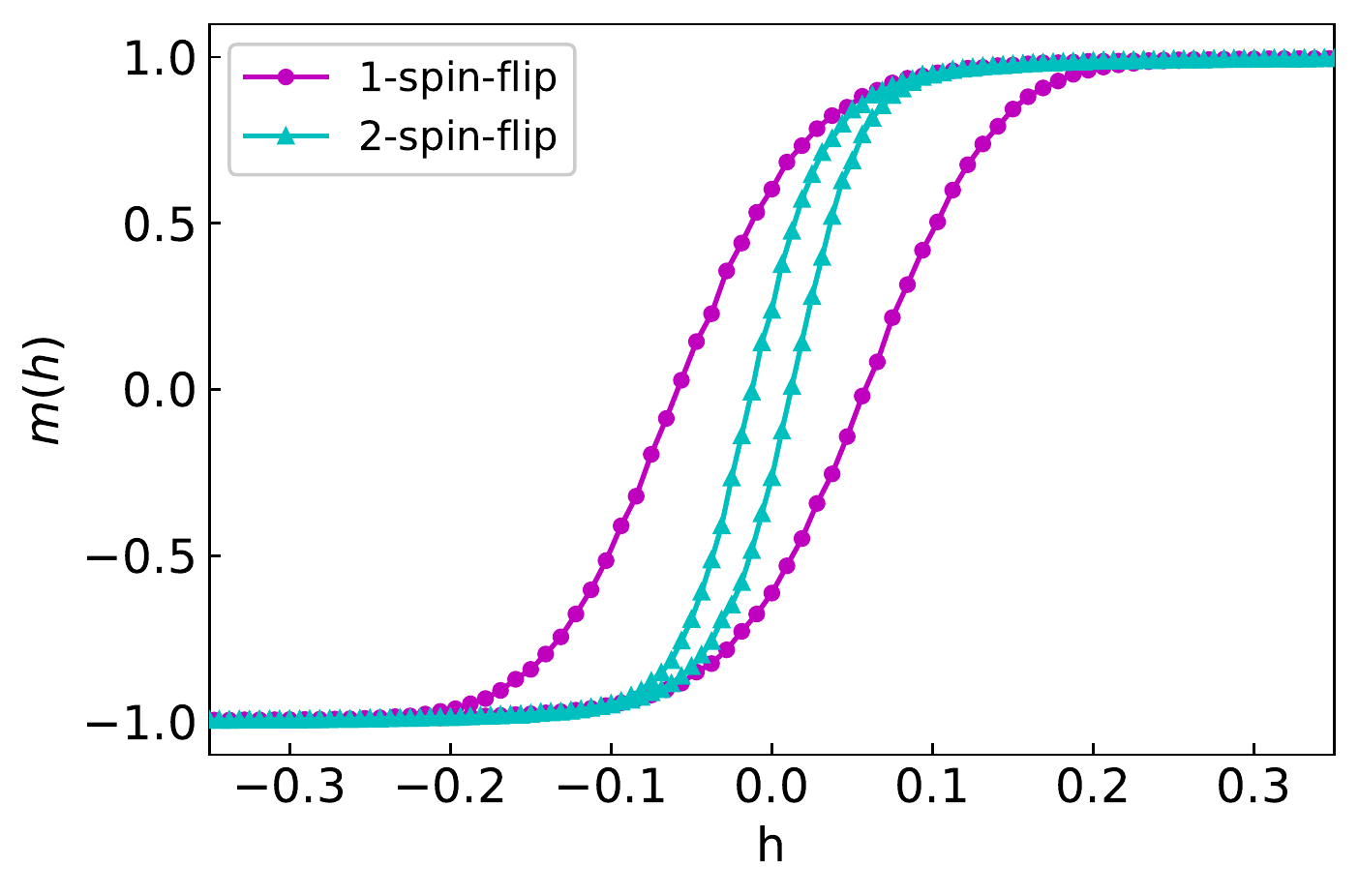}
   \caption{(Color online) Figure shows the smaller area of hysteresis 
   loops on implementing 2-spin-flip dynamics compared to 1-spin-flip 
   for the same set of parameters $L=2000$, $K=1.5$ and $t=50$. All data 
   are averaged over 100 configurations.}
  \label{fig:3}
\end{figure}  

\subsection{hysteresis on regular random graphs}

\begin{figure}[htb]
 \centering
\includegraphics[width=0.45\textwidth,angle=0]{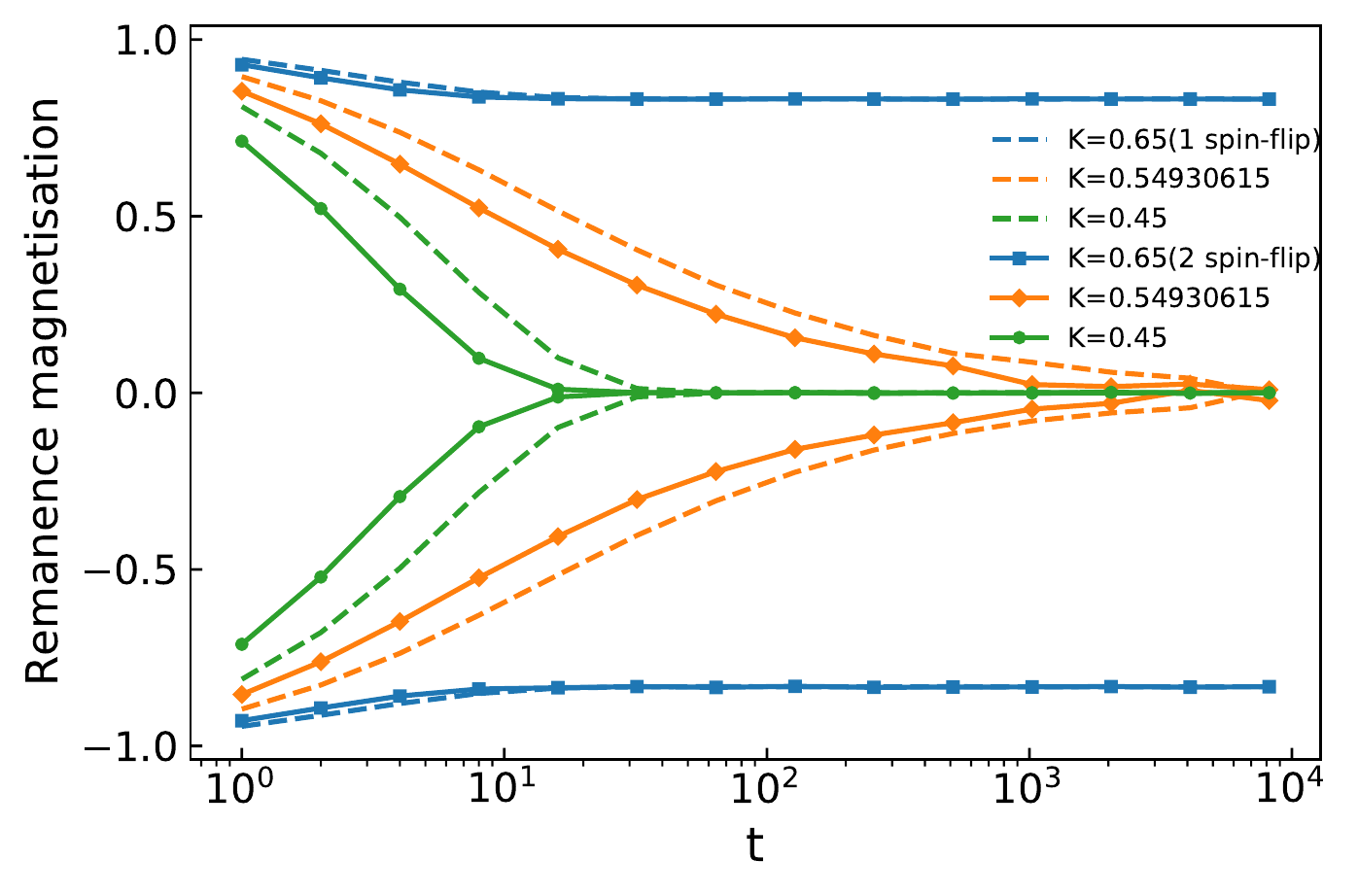} 
   \caption{(Color online) The upper half of the figure shows how 
   initial saturated magnetization in the up direction decays under 
   various relaxation algorithms at zero applied external field and 
   different temperatures. Similarly the lower half shows the increase 
   of magnetization with time under relaxation on $z=3$ random graph for $L=10^6$ (averaged over 20 
   configurations). Other details as indicated in the figure.}

\label{fig:4}
\end{figure}  

A more promising route to explore the role of relaxation dynamics on the 
nature of equilibrium state which is eventually obtained is to go beyond 
a one dimensional Ising model. The one dimensional Ising model does not 
have a finite critical temperature. In other words, it does not order 
spontaneously below a finite temperature $T_c$. The simplest model which 
has a finite critical temperature $T_c$ whose value can be obtained 
analytically is the Ising model on a Bethe lattice with coordination 
number $z>2$. In this case the inverse critical temperature is given by 
the equation $\tanh K_c = 1/(z-1)$~\cite{rozikov}. For $z=3$, we get 
$K_c=0.5493$ approximately. The theory on the Bethe lattice can be 
checked quite effectively by simulations on regular random graphs. It is 
difficult to pin point the critical temperature exactly using dynamics 
but the presence of a watershed between high $K$ and low $K$ behaviour 
is easily spotted. Initial studies suggest the critical value $K_c$ 
obtained from 1-flip dynamics is in close vicinity of the analytical 
value $K_c=0.5493$ obtained from the equilibrium partition function. 
There may be some reason to expect $K_c$ obtained by dynamics to 
increase slightly as we go from one-flip to two-flip dynamics. This is 
because a combination of one and two spin flips may be able to explore 
deeper energy minima of the system. However we do not see a clear 
indication of this effect in our numerical work so far. Fig.4 presents 
the results of simulations at external applied field $h=0$, and three 
different values of $K=0.65, 0.5493,$ and $0.45$. At each value of $K$, 
two sets of magnetization trajectories are shown in the same color, one 
starting with all spins down and the other with all spins up initially. 
At $K=K_c$ the magnetization approaches zero relatively slowly 
irrespective of the initial state. For $K < K_c$, it does the same but 
much faster as may be expected. For $K > K_c$, the system relaxes very 
slowly as may be expected and the magnetization remains close to the 
initial value over much of the observed times. We examined more closely 
the cases $K \le K_c$ in the long time limit. Here the magnetization 
reaches a value nearly equal to zero within statistical fluctuations and 
the system may be said to be in a state of equilibrium. We examined if 
the fraction of neighboring spins parallel to reach other were in a 
higher proportion with two spin flips than with one spin flip dynamics. 
But we did not find any significant difference.

\section{Future prospects}

Our main concern in this article has been to investigate the equilibrium 
state of an isolated thermodynamic system. As Feynman ~\cite{feynman} 
put it, equilibrium is when fast things have happened and slow things 
have not. Thus equilibrium necessarily implies a separation between 
short and long time scales. Let us say $\Gamma$ is the characteristic 
time that separates these two. Then a system in equilibrium is 
thermalised with respect to fast relaxation processes with characterises 
time $\gamma << \Gamma$ but not for slower processes. A great fortune of 
equilibrium statistical mechanics is that in the limit $\Gamma 
\rightarrow \infty$, properties of a system can be deduced from a 
knowledge of its partition function without knowing the dynamics. 
However a misfortune is that we can hardly ever evaluate a partition 
function exactly. Also experimental observations are obtained 
necessarily at a finite $\Gamma$. Thus we must resort to solving the 
dynamical equations which are not only too complex and coupled but also 
not known exactly for an extended complex system. We have to work with a 
set of model equations that hopefully might be tractable. Some hope in 
this direction is provided by a few models whose partition function or 
critical temperature has been evaluated exactly~\cite{wannier}. These are Ising models 
in one and two dimensions and on Bethe lattices of a general 
coordination number $z$. The Hamiltonians of these models are not true 
Hamiltonians in the sense of classical mechanics. The so called 
Hamiltonian does not generate Hamilton's equations of motion. We have to 
put in a stochastic dynamics like Glauber's dynamics by hand. Such 
dynamics can have different variants, e.g. single-flips or two-flips 
etc. There is no reason to expect if any of these variants would yield 
the same equilibrium state as the partition function. This is indeed a 
conundrum. On one hand we are accustomed to the belief that dynamics is 
unimportant in determining the properties of a system in thermal 
equilibrium. On the other hand it does look reasonable that the state of 
the system after a long time may retain some signature of the dynamics 
that brought it there. The preliminary work presented above is 
unfortunately not very conclusive in resolving this issue. One and two 
spin flips both appear to lead to equilibrium states that are hard to 
distinguish from each other. This may be an artifact of the Bethe 
lattice or may indicate a deeper truth of statistical mechanics. 
Nonetheless in our opinion it does raise an issue which has not received 
sufficient attention so far and needs to be explored further. We hope it 
is a small first step on a fruitful path.

\section{Acknowledgement}

This write up has evolved from an earlier published work 
~\cite{shukla1}, and online presentations ~\cite{shukla2,thongjaomayum} 
in conferences over the last year or so.


\begin{thebibliography}{99}

\bibitem{wilson} See for example, K G Wilson and J Kogut, Phys. Rep. 
C 12, 77 (1974), and references therein.

\bibitem{ising} E Ising, Z Phys 31, 253 (1925).

\bibitem{onsager} L Onsager, Phys Rev 65, 117 (1944).

\bibitem{glauber} R J Glauber, J Math Phys 4, 294 (1963).

\bibitem{bertotti} See for example {\em{The Science of Hysteresis}} 
edited by G Bertotti and I Mayergoyz (Academic Press, Amsterdam, 2006).


\bibitem{rozikov} U A Rozikov, {\em{Gibbs Measures on Cayley Trees}}, 
World Scientific (2013). 

\bibitem{feynman} Statistical Mechanics: A set of lectures, by Richard P 
Feynman, Taylor and Francis, ABC ppbk, ISBN0-201-36076-2 (2018).

\bibitem{wannier} G H Wannier, Rev Mod Phys 17, 50 (1945); Phys Rev 79, 
357 (1950).

\bibitem{shukla1} Prabodh Shukla, Phys Rev E97, 062127 (2018).

\bibitem{shukla2} Prabodh Shukla in Statistical Physics: Recent Advances 
and Future Directions (Online), ICTS, Bangalore (15 Feb, 2022).

\bibitem{thongjaomayum} Diana Thongjaomayum, XIII Biennial National Conference of Physics Academy of North East (2022), Manipur University, Imphal; $2^{nd}$ International Conference on Advancement in Core and Frontier of Physics (2023), GLA University, Mathura.


\end{thebibliography}
\end{document}